\documentstyle[12pt]{article}

\def\be{\begin{equation}}
\def\ee{\end{equation}}
\def\beq{\begin{eqnarray}}
\def\eeq{\end{eqnarray}}
\def\bes{\begin{subequation}}
\def\ees{\end{subequation}}
\def\beseq{\begin{subeqnarray}}
\def\eeseq{\end{subeqnarray}}

\catcode`@=12
\begin{document}
\thispagestyle{empty}
\begin{flushright}
TIFR/TH/97-12\\
UCRHEP-T186\\
(revised)
\end{flushright}
\bigskip

\begin{center}
{\large{\bf New interactions in neutrino oscillations with three light
flavors}} \\[2cm]
Ernest Ma\footnote{email: ernestma@ucrac1.ucr.edu} \\
Physics Department \\
University of California, Riverside \\
CA 92521. \\[.5cm]
Probir Roy\footnote{email: probir@theory.tifr.res.in} \\
Tata Institute of Fundamental Research \\
Mumbai 400 005, India. \\[1cm]
September 1997 \\[1cm]
Abstract
\end{center}
\bigskip

If one assumes solar and LSND neutrino oscillations to explain the
corresponding data, then the atmospheric neutrino deficit cannot be 
accommodated within the Standard Model with three light flavors, unless 
one ignores the data's zenith-angle dependence.  We propose a novel 
solution to this problem by postulating large anomalous diagonal
$\nu_\tau$-quark interactions which affect $\nu_\mu - \nu_\tau$ 
oscillations traversing the Earth and induce the observed 
zenith-angle dependence.

\newpage
\baselineskip 24pt

Three flavors of massless lefthanded weakly interacting neutrinos
occur in the Standard Model.  Experimental studies at the
Large Electron Positron (LEP) collider have definitively established
the number of light weak neutrino flavors to be three.  However, the
masslessness of any neutrino is not predicted on fundamental grounds.
Beyond the Standard Model, theoretical arguments exist showing how
neutrinos could acquire tiny Majorana [1] or Dirac [2] masses.  On the
experimental front, there is indirect evidence of small nonvanishing
neutrino masses from three different kinds of phenomena
pertaining to neutrino oscillations.  (1) The observed depletion [3]
of the solar neutrino flux from the prediction of the standard solar
model in different segments of the solar neutrino energy spectrum, (2)
the claimed discovery [4] of $\bar\nu_\mu - \bar\nu_e$ and $\nu_\mu - \nu_e$ 
oscillations by the Liquid Scintillator Neutrino 
Detector (LSND) experiment and (3) the deficit [5] in the atmospheric
neutrino flux, measured on the ground in terms of the ratio of ratios
$R \equiv (\nu_\mu + \bar\nu_\mu):(\nu_e + \bar\nu_e)_{\rm expt.}/(\nu_\mu
+ \bar\nu_\mu):(\nu_e + \bar\nu_e)_{MC}$, $MC$ standing for the
Monte-Carlo expectation -- all point
to nonzero neutrino masses. 
The canonical best fits to the data from the above three different
experimental studies cannot be accommodated within the hypothesis of 
only three light neutrino flavors because of the three nonoverlapping 
ranges of $\delta m^2$ involved.  This and other considerations from 
astrophysics and cosmology have led to the speculation of
the existence [6] of a fourth sterile (i.e. electroweak singlet) light
neutrino as a possible way of reconciling all of the known data.  On the 
other hand, if only three neutrino flavors are assumed, two of the three 
possible neutrino oscillations can be explained and the question is to 
what extent the third can be accommodated.  Previous studies have chosen 
the exception to be either the atmospheric data [7] or the solar data [8]. 
A recent general analysis [9] shows that the former hypothesis is in fact 
favored.

It seems to us that the results of the solar neutrino and the LSND
experiments are quite unambiguous, assuming the absence of unknown
sources of systematic error.  In contrast, the detailed conclusions
from the atmospheric neutrino experiment seem to depend sensitively on
the intricacies of the Monte-Carlo simulations [5] used.  It may,
therefore, be more profitable to consider three-flavor scenarios which
naturally explain the solar neutrino and the LSND data and then to
explore the observed atmospheric neutrino anomaly within those.  That
leads one naturally to the scheme of Ref.~[7].  This scheme, in which
the interactions are exclusively those of the Standard Model, leads to
a universal value of $R$ that can give acceptable fits to the sub-GeV and
multi-GeV data, integrating over all zenith angles. However, it
disallows any measurable dependence of the data on the zenith angle
predicting an essentially flat distribution.  

As in Ref.~[7], we too start with only three light neutrino flavors and
Standard Model interactions, but then we extend the latter to include
the possibility of anomalous neutrino interactions [10].
Specifically, we allow $\nu_\tau$ to have any large nonstandard
diagonal four-fermion (effectively contact) interaction with quarks.
This is motivated by two facts: (1) the reported observation [11] of
anomalous $e^+$-quark interactions at HERA, which suggests the possibility 
of anomalous lepton-quark interactions in general; (2) among the three known
neutrinos, $\nu_{\mu,e}$ are strongly forbidden by experimental
constraints to have such large interactions while there exist
essentially no restrictions on $\nu_\tau$.  (In particular, there
could be a heavy vector boson coupling only to leptons of the third
generation but to the light quarks as well).  As we show below, this will 
result in the \underline {novel possibility} of an induced zenith-angle 
dependence for 
atmospheric neutrino oscillations, which appears to be favored by the data.

With an anomalous $\nu_\tau$-quark interaction, the survival probability 
$R$ may vary with the zenith
angle in atmospheric neutrino oscillations despite a large $\delta m^2$ 
chosen at around 0..25 eV$^2$ to satisfy the LSND data.  Since the 
interaction cross section (in the detector) of neutrinos is roughly 
3 times that of antineutrinos, and the neutrino flux is somewhat larger 
than the antineutrino flux [12] at higher energies, such a variation 
is potentially able to explain the multi-GeV 
atmospheric data.  The extra $\nu_\tau$ interactions inside the sun are 
offset by a $\delta m^2$ much larger 
than it would be for the canonical matter-enhanced effect, as explained below.

We start with the following approximate mass eigenstates: 
$\nu_1 \sim \nu_e$,  $\nu_2 \sim c_0 \nu_\mu + s_0 \nu_\tau$, 
$\nu_3 \sim -s_0 \nu_\mu + c_0 \nu_\tau$,
where $c_0 \equiv \cos \theta_0$, $s_0 \equiv \sin \theta_0$, and $\theta_0$ 
is not small.  We choose $m_1 \sim 0$, $m_2 \sim 10^{-2}$ eV, and 
$m_3 \sim 0.5$ eV.  We then allow $\nu_1$ to mix with $\nu_3$ with a small 
angle $\theta'$ and the new $\nu_1$ to mix with $\nu_2$ with a small angle 
$\theta$.  The exact mass eigenstates are then
\begin{eqnarray}
\nu_1 &=& c c' \nu_e + c s' (-s_0 \nu_\mu + c_0 \nu_\tau) - s (c_0 \nu_\mu 
+ s_0 \nu_\tau), \\ \nu_2 &=& c (c_0 \nu_\mu + s_0 \nu_\tau) + s c' \nu_e 
+ s s' (-s_0 \nu_\mu + c_0 \nu_\tau), \\ \nu_3 &=& c' (-s_0 \nu_\mu + c_0 
\nu_\tau) - s' \nu_e.
\end{eqnarray}

Let us now turn one-by-one to the three sets of neutrino oscillation
data. 

\noindent {\bf LSND}:~~
$\bar \nu_\mu - \bar \nu_e$ oscillations, as probed in this experiment, are 
controlled by $\delta m_{31}^2 \sim 0.25$ eV$^2$.  Any significantly higher 
value chosen for $\delta m_{31}^2$ will be in contradiction with restrictions 
imposed by the search for $\nu_\mu$ disappearance in the CDHS experiment [13] 
for large angles (which will be needed later in explaining the atmospheric 
neutrino effect).  On the other hand, for such a value of $\delta m^2$, the 
LSND data [4] imply a mixing angle $\chi$ with $\sin 2 \chi \simeq 0.19$. 
These numbers are just about compatible with the constraints of the Bugey 
experiment [14].  Comparing with (3), we find
$2s_0 s'c' \simeq 0.19$.

\noindent {\bf Solar neutrino data}:~~
The canonical solution for solar neutrino oscillations takes $\nu_1 \sim 
\nu_e$ and $\nu_2 \sim$ a linear combination of $\nu_\mu$ and $\nu_\tau$, 
with $m_2 > m_1$ and some mixing between $\nu_1$ and $\nu_2$.  In its 
passage through the sun, $\nu_e$ gets an extra induced mass because of its 
forward scattering with the electrons.  The matching of this mass with 
$\delta m_{21}^2$ produces the well-known MSW effect [15].  Here we have 
new extra diagonal $\nu_\tau$-quark interactions.  Consequently, a larger 
$\delta m_{21}^2$ is needed to cancel against the induced $\nu_\tau$ mass, 
which should be negative in this case [16].  

Let us examine in detail the effect of $\nu_\tau$-quark interactions on the
passage of electron neutrinos through the sun.  We write this new 
interaction as [10]
\be
{\cal L}_{\rm new} = -\sqrt{2} \bar\nu_{\tau L} \gamma_\mu \nu_{\tau
L} (G^q_{\tau\tau V} \bar q \gamma^\mu q + G^q_{\tau\tau A} \bar q
\gamma^\mu \gamma_5 q)
\ee
for all quarks $q$.  Note that the $\tau$ neutrino and the
$\tau$ antineutrino are known  
from $\tau$-decay properties to be lefthanded and righthanded
respectively.  Therefore, all possible four-fermion interactions
involving them and quarks can be brought into the form (4) by Fierz
transformations.  Only the vector coupling $G^q_{\tau\tau V}$
contributes to the potential relevant to forward scattering while its
contributions for the neutrino and antineutrino cases are equal in
magnitude but opposite in sign.  We define 
$\epsilon'_q \equiv G^q_{\tau\tau V}/G_F$
as in Ref.~[10].  Note that we do not require any 
flavor-changing interactions which would have been necessary to obtain 
oscillations if $\delta m^2 = 0$.

Solar neutrino oscillations occur between $\nu_e$ and $\nu_\alpha = 
c_0\nu_\mu + s_0\nu_\tau$ with angle $\theta$.  Since $m^2_3 \gg m_2^2$, 
$\nu_\beta = -s_0\nu_\mu + c_0\nu_\tau$ effectively decouples. Hence the 
relevant evolution equation can be written, after rotating away a common 
phase, as [10]
\be
4i E_\nu {d \over dt} \left(\matrix{\nu_e \cr \nu_\alpha}\right)
\simeq \left(\matrix{0 & m^2_2 \sin 2\theta \cr m^2_2 \sin
2\theta & 2m^2_2 \cos 2\theta + 4\sqrt{2} G_F E_\nu (s_0^2
\epsilon'_q N_q - N_e)}\right) \left(\matrix{\nu_e \cr
\nu_\alpha}\right).
\ee
In (5), $N_q \epsilon'_q \equiv N_u \epsilon'_u + N_d \epsilon'_d$
and $N_{e,u,d}$ is the number of (electrons, $u$-quarks, $d$-quarks)
per unit solar volume.  The coefficient $s_0^2 \equiv \sin^2 \theta_0$ of
the $\epsilon'_q N_q$ term in the second diagonal matrix element
originates from the $3 \times 3 \rightarrow 2 \times 2$ flavor matrix
reduction.  In order to have a large $\epsilon'_q$ and yet satisfy the 
resonance condition for solar-neutrino flavor conversion, we see that 
$m_2$ should be larger than its canonical value of $2.45 \times 10^{-3}$ eV, 
and $\epsilon'_q$ should be negative.  [If $\epsilon'_q$ comes from $R$-parity 
violating squark exchange, then it must be positive; but if it comes from 
vector exchange, then it may be of either sign.]

We now assume as a crude approximation that $N_q \simeq 4 N_e$ in the sun.  
It then follows from (5) that the effective mixing angle for
$\nu_e - \nu_\alpha$ oscillations in solar matter is
given by 
\be
\tan 2\theta^S_m = {\sin 2\theta \over \cos 2\theta + 2\sqrt{2}
G_F m^{-2}_2 E_\nu (4s_0^2 \epsilon'_q - 1) N_e}
\ee
and the MSW resonance condition [15] is
\be
m^{-2}_2 (-4s_0^2 \epsilon'_q + 1) = \cos 2\theta (2\sqrt{2} G_F N_e
E_\nu)^{-1}. 
\ee
In the canonical MSW solution, the left-hand side is $(6 \times 10^{-6} 
{\rm eV}^2)^{-1}$.  It can thus be matched with the requirement
\be
s_0^2 \epsilon'_q \simeq -3.92 = -4.17 (m_2^2/10^{-4} {\rm eV}^2) + 0.25.
\ee
Our seemingly arbitrary 
choice of $\delta m^2_{21} \sim 10^{-4}~{\rm eV}^2$ is now seen as a 
reasonable value so that $\epsilon'_q$ can be large enough to be relevant 
for the following discussion on the atmospheric neutrino data.  Note that 
the same range of $\theta$, {\it i.e.} near 0.04, works here as well as 
in the standard MSW solution for solar neutrino oscillations. 

\noindent {\bf Atmospheric neutrino data}:~~
The depletion in the flux of muon neutrinos and antineutrinos, produced in the 
earth's upper atmosphere, is caused by $\nu_\mu - \nu_\tau$ flavor 
oscillations which occur between the physical states $\nu_2$ and 
$\nu_3$ with $\delta m_{32}^2 \sim 0.25$ eV$^2$ and angle $\theta_0$.  
The oscillation wavelength ``in vacuo" is $\lambda \sim 
4 \pi E_\nu/\delta m_{32}^2 \sim 10 (E_\nu$/GeV) km, $E_\nu$ being the 
neutrino energy.  Hence, for $E_\nu < 10$ GeV, 
several oscillations occur in the earth's atmosphere.  Consequently, one 
obtains the classical survival probability $P_0 = c_0^4 + s_0^4 = 1 - 
{1 \over 2} \sin^2 2 \theta_0$.  The choice of $s_0 \simeq 0.47$
yields $R = P_0 \simeq 0.66$ for the ratio of ratios, in reasonable 
agreement with the data for downward going neutrinos and antineutrinos. 
This implies $s' \simeq 0.21$.

For neutrinos and antineutrinos coming downward, the density of the atmosphere 
is negligible for the new diagonal $\nu_\tau$-quark interactions to be of 
any importance.  For upward moving ones, the density of the earth turns 
out to be in the right range for them to make a difference.  
Since $\nu_e$ gets effectively decoupled from the $\nu_\mu -
\nu_\tau$ oscillation problem,  we now have 
\be
4i E_\nu {d \over dt} \left( \matrix {\nu_\mu \cr \nu_\tau}\right) \simeq 
\left(\matrix{0 & m^2_3\sin 2\theta_0
\cr m^2_3\sin 2\theta_0 & 2m^2_3\cos 2\theta_0 +
4\sqrt{2} G_F E_\nu \epsilon'_q N_q}\right) \left(\matrix{\nu_\mu \cr
\nu_\tau}\right). 
\ee

For the earth we estimate an average $N_q \sim 9 \times 10^{30}$ 
m$^{-3}$.  Thus if one chooses to define the parameter 
$X \equiv \epsilon'_q E_\nu/(10 \ {\rm GeV})$, 
the effective mixing angles in
terrestrial matter (a) $\theta^E_m$ between $\nu_\mu$ and $\nu_\tau$
and (b) $\bar\theta^E_m$ between $\bar\nu_\mu$ and $\bar\nu_\tau$ are
respectively given by
\be
\tan 2\theta^E_m = {\sin 2\theta_0 \over \cos 2\theta_0 + 0.091 X}, ~~~ 
\tan 2\bar\theta^E_m = {\sin 2\theta_0 \over \cos 2\theta_0 - 0.091 X},
\ee
with $\cos 2\theta_0 = \mp 0.091 X$ as the resonance conditons.

For sub-GeV neutrinos, the $X$ term is insignificant, but 
for multi-GeV neutrinos it may become large enough for the 
resonance condition to be satisfied. 
Assuming adiabaticity, the neutrino and antineutrino 
flavor survival probabilities are described well by the formulae [17]
\beq
P(\nu_\mu \rightarrow \nu_\mu) &=& {1\over2} (1 + \cos 2\theta_0 \cos
2\theta^E_m), \\[2mm]
\bar P(\bar\nu_\mu \rightarrow \bar\nu_\mu) &=& {1\over2} (1 + \cos
2\theta_0 \cos 2\bar\theta^E_m),
\eeq
where $\cos 2\theta^E_m$ and $\cos 2\bar\theta^E_m$ are computed from
(11).  Although the conditions for adiabaticity 
may not be satisfied, our purpose is to try to find the maximum effect 
for a given magnitude of the anomalous interaction which is of course 
unknown.  Any nonadiabaticity would only tend to diminish this effect.  

Owing to the opposite signs of the media contributions to neutrino
and antineutrino oscillations, matter effects will get somewhat 
diluted.  However, there are two important 
factors to be considered. First, the initial $\nu_\mu$ 
flux is larger than the $\bar \nu_\mu$ flux for multi-GeV neutrinos.  
In the upper atmosphere, $\nu_\mu (\bar \nu_\mu)$ is produced together 
with $\mu^+ (\mu^-)$ from $\pi^+ (\pi^-)$ decay.  The subsequent decay 
of $\mu^+ (\mu^-)$ to $\bar \nu_\mu (\nu_\mu)$ will equalize the total 
number of $\nu_\mu$ and $\bar \nu_\mu$, but there is an energy dependence 
and given that the $\mu^+/ \mu^-$ ratio is larger than one [12], we allow 
a factor of $r$ = ratio of the $\nu_\mu$ to $\bar \nu_\mu$ flux in our 
following discussion.  Second, in the detection of atmospheric neutrinos, 
there is no experimental measurement of the charge of the resulting leptons. 
Specifically, $\mu$-like events include both $\mu^-$ and $\mu^+$, but 
they are not separated.  Now $\sigma_\nu \simeq 3 \sigma_{\bar \nu}$ for 
an isoscalar target, hence the measured probability is weighted:
\begin{equation}
P_m \simeq {{3 r P + \bar P} \over {3 r + 1}}.
\end{equation}
In the absence of media effects, $P = \bar P$ even with oscillations 
(assuming CP conservation), hence $P_m = P = \bar P$.  In the presence 
of media effects, $P \neq \bar P$, so the above expression for $P_m$ 
should be used.

Suppose we now make the same choice of $s_0 \simeq 0.47$
which leads to a suppression probability $P_0 = 0.66$ for the downward
travelling neutrinos.  Then, for an optimistically large and negative value of
$X = -15$, we get $P = 0.31$ and $\bar P = 0.76$, hence $P_m$ is lowered  
to 0.39 if $r = 1.5$ in Eq.~(14) or 0.42 if $r = 1.0$.  This 
is a potential explanation of the observation of a smaller $R$ for 
atmospheric neutrinos going through the earth (zenith angle $> \pi/2$) 
where $R = P_m$ than for those coming down through only the atmosphere
(zenith angle $< \pi/2$) where $R = P_0$.

There have been studies [18] of $R$-parity violating 
squark interactions, scalar and vector leptoquarks, as well as contact 
interactions of neutrinos.  Whereas these may be related to our 
proposed effective interaction, they are all restricted to be small. 
We consider instead a vector boson $B$ which couples to
$\bar u \gamma^\mu u + \bar d \gamma^\mu d - \bar \nu_\tau \gamma^\mu \nu_\tau 
- \bar \tau_L \gamma^\mu \tau_L$
with coupling $g_B$.  This interaction would result in a negative $X$ as 
desired.  Using the result of a previous model [19] where $B$ 
couples to baryon number, and allowing for the fact that here $B$ couples 
to $u$ and $d$ but not $s$, $c$, and $b$, we find
$\alpha_B \equiv g_B^2/4\pi \leq 0.057$.
We now require $g_B^2/m_B^2 = 15 G_F$, hence $m_B = 64$ GeV is allowed.  The 
deviation from $e - \mu - \tau$ universality in $Z$ decay is then given 
by
\begin{equation}
{\alpha_B \over 2 \pi} \left( {g_L^2 \over g_L^2 + g_R^2} \right)_\tau 
(F_1 + F_2) = {0.057 \over 2 \pi} (0.5735) (1.32) = 0.0069,
\end{equation}
where $F_1$ and $F_2$ are well-known functions [19] of the ratio $m_B^2/M_Z^2$ 
which is about 0.5 for $m_B = 64$ GeV.  The above is exactly two standard 
deviations from the experimental data [20], taking into account the 
kinematical correction due to $m_\tau$ in $Z \to \tau^+ \tau^-$.  The 
deviation in the total invisible width of $Z$ is 0.012 versus the 
standard-model value of 3, which is again two standard deviations from the 
data [20].  This is then a possible explicit model for our scenario. 

Our scenario will have the following consequences
for the forthcoming experiments.  Solar neutrino experiments will confirm 
the MSW solution, but the interpretation of $\delta m^2$ is subject to the 
ambiguity that it could be $\delta m^2 /(1 - 4 s_0^2 \epsilon'_q)$ instead. 
However, the anomalous $\nu_\tau$-quark interactions will be observable 
at SNO, thereby resolving this ambiguity. 
Both $\nu_\mu \rightarrow \nu_e$ and $\nu_\mu \rightarrow \nu_\tau$ 
conversion experiments will measure a $\delta m^2$ at around 0.25 eV$^2$, 
but $\sin^2 2 \theta_{\mu,e} \sim 0.036$ while $\sin^2 2 \theta_{\mu,\tau} 
\sim 0.69$.  The former is outside the region being probed by reactor 
experiments ($\bar \nu_e$ disappearance) such as Chooz and Palo Verde; 
the latter is outside that being probed by short-baseline accelerator 
experiments ($\nu_\tau$ appearance) such as CHORUS and NOMAD.  On the 
other hand, both regions are covered by all the proposed long-baseline 
experiments (either through $\nu_\mu$ disappearance, or $\nu_e$ and 
$\nu_\tau$ appearances) such as MINOS, K2K (KEK-PS/Super-Kamiokande) and 
CERN-SPS/ICARUS.

More immediately, the new data from Super-Kamiokande, Soudan 2, and 
MACRO on $\nu_\mu + \bar \nu_\mu$ events through the earth will be sensitive 
to the anomalous $\nu_\tau$-quark interactions.  There should be an energy 
dependence as well as a zenith-angle dependence.  In particular, the 
zenith-angle dependence should be absent or much smaller for sub-GeV data. 
For a zenith angle near zero, our proposal is easily distinguishable from the 
$\delta m^2 \sim 10^{-2}$ eV$^2$ oscillation interpretation because we have 
$R = P_0$ whereas the latter would require $R \sim 1$, owing to the short 
distance between production and detection in that case.  
To test our hypothesis further, the detection and acceptance efficiencies 
of neutrinos versus antineutrinos have to be understood in more detail. 
Better yet, the capability of these experiments for distinguishing neutrinos 
from antineutrinos should be explored [21].  
\medskip

\noindent {\it Acknowledgements}

P.R. wishes to acknowledge the hospitality of the Department of
Physics, UC Riverside, during the course of this work.  This work was 
supported in part by the U. S. Department of Energy under Grant No. 
DE-FG03-94ER40837.

\newpage
\begin{center}
{\bf REFERENCES}
\end{center}
\bigskip

\begin{enumerate}
\item[{[1]}] M. Gell-Mann, P. Ramond and R. Slansky, in {\it
Supergravity} (ed. P. Van Nieuwenhuizen and D.Z. Freedman,
North-Holland, 1979), p315.  T. Yanagida, in Proceedings of the
Workshop on the Unified Theory and the Baryon Number in the Universe
(ed. O. Sawada and A. Sugamoto, KEK Report No. 79-18, Tsukuba, Japan,
1979). 
\item[{[2]}] P. Roy and O. Shanker, Phys. Rev. Lett. {\bf 52}, 713
(1984); Phys. Rev. {\bf D30}, 1949 (1984); M. Roncadelli and D. Wyler,
Phys. Lett. {\bf 133B}, 325 (1983).
\item[{[3]}] For reviews of the subject and many references see the
following articles: R.S. Raghavan, Science {\bf 267}, 45 (1995).
J.H. Bahcall and M.H. Pinsonneault, Rev. Mod. Phys. {\bf 67}, 1
(1995).  R. Davis, Prog. Part. Nucl. Phys. {\bf 32}, 13 (1994).
Y. Fukuda et al., Phys. Rev. Lett. {\bf 77}, 1683 (1996).
P. Anselmann et al., Phys. Lett. {\bf B327}, 377 (1994); {\bf 342},
440 (1995).  J.N. Abdurashitov et al., Phys. Lett. {\bf B328}, 234
(1994). 
\item[{[4]}] C. Athanassopoulos et al., Phys. Rev. Lett. {\bf 75},
2650 (1995); {\bf 77}, 3082 (1996).  The high $\delta m^2$ region,
allowed by this experiment, has recently been ruled out by the CCFR
group: A. Romosan et al., Phys. Rev. Lett. {\bf 78}, 2912 (1997). The
observation of $\nu_\mu \leftrightarrow \nu_e$ oscillations has been
recently claimed by the LSND group: C. Athanassopoulos et al., 
nucl-ex/9706006.
\item[{[5]}] Y. Fukuda et al., Phys. Lett. {\bf B335}, 237 (1994) and
earlier references therein.  R. Becker-Szendy et al., Phys. Rev. {\bf
D46}, 3720 (1992).  D. Casper et al., Phys. Rev. Lett. {\bf 66}, 2561
(1991). 
\item[{[6]}] J.R. Primack, J. Holtzman, A. Klypin and D.O. Caldwell,
Phys. Rev. Lett. {\bf 74}, 2160 (1995).  G.M. Fuller, J.R. Primack, and Y.-Z. 
Qian, Phys. Rev. {\bf D52}, 1288 (1995).. E. Ma and P. Roy, Phys. Rev. {\bf
D52}, R4780 (1995).  See also D.O. Caldwell and R.N. Mohapatra, Phys. Rev. 
{\bf D48}, 3259 (1993); J.T. Peltoniemi and J.W.F. Valle, Nucl. Phys. 
{\bf B406}, 409 (1993).
\item[{[7]}] C. Y. Cardall and G. M. Fuller, Phys. Rev. {\bf D53}, 4421 
(1996).  See also the earlier work of S. M. Bilenky, A. Bottino, C. Giunti, 
and C. W. Kim, Phys. Lett. {\bf B356}, 273 (1995);  K. S. Babu, J. C. Pati, 
and F. Wilczek, {\em ibid.}, {\bf B359}, 351 (1995); G. L. Fogli, E. Lisi, 
and G. Scioscia, Phys. Rev. {\bf D52}, 5334 (1995).
\item[{[8]}] A. Acker and S. Pakvasa, Phys. Lett. {\bf B397}, 209 (1997).
However, the prediction of energy independence here is 
disfavored by the data: P.I. Krastev and
S.T. Petcov, Phys. Lett. {\bf B395}, 69 (1997).
\item[{[9]}] G.L. Fogli, E. Lisi, D. Montanino, and G. Scioscia, Phys. Rev. 
{\bf D56}, 4365 (1997).
\item[{[10]}] M.M. Guzzo, A. Masiero, and S.T. Petcov, Phys. Lett. {\bf B260}, 
154 (1991); E. Roulet, Phys. Rev. {\bf D44}, R935 (1991); V. Barger, R.J.N. 
Phillips, and K. Whisnant, Phys. Rev. {\bf D44}, 1629 (1991); P.I. Krastev 
and J.N. Bahcall, hep-ph/9703267.
\item[{[11]}] H1 Collaboration, C. Adloff {\it et al.}, Z. Phys. {\bf C74}, 
191 (1997);  ZEUS Collaboration, J. Breitweg {\it et al.}, Z. Phys. 
{\bf C74}, 207 (1977).
\item[{[12]}] Preliminary results of the HEAT and MASS cosmic ray experiments 
on the ratio $\mu^+/\mu^-$ are $1.64 \pm 0.08$ and $1.2 \pm 0.05$ 
respectively, as quoted by M. Goodman in {\it Long Baseline Neutrino 
Newsletter} (August 1997).
\item[{[13]}] F. Dydak et al., Phys. Lett. {\bf B134}, 281 (1984).  This 
$\nu_\mu$ disappearance experiment by itself is usually not taken to 
constrain the LSND data which concern only $\nu_\mu \rightarrow \nu_e$, but 
in our case, it is a relevant constraint because we will also consider 
$\nu_\mu \rightarrow \nu_\tau$ as a possible explanation of the atmospheric 
neutrino anomaly, using the same $\delta m^2 \sim 0.25 ~{\rm eV}^2$.
\item[{[14]}] Y. Declais et al., Nucl. Phys. {\bf B434}, 503 (1995).
\item[{[15]}] L. Wolfenstein, Phys. Rev. {\bf D17}, 2369 (1978).
S. Mikheyev and A.Yu. Smirnov, Sov. J. Nucl. Phys. {\bf 42}, 913
(1985). 
\item[{[16]}] If it is positive, then an inverted mass hierarchy, {\it i.e.} 
$m_2 < m_1$ would be needed.
\item[{[17]}] T. Kuo and J. Pantaleone, Rev. Mod. Phys. {\bf 61}, 937
(1989). 
\item[{[18]}] G. Bhattacharyya {\em et al.}, Mod. Phys. Lett. {\bf A10}, 
1583 (1995); J. K. Mizukoshi {\em et al.}, Nucl. Phys. {\bf B443}, 20 (1995); 
O. Eboli {\em et al.}, Phys. Lett. {\bf B396}, 238 (1997); M. Bilenkii and 
A. Santamaria, Phys. Lett. {\bf B336}, 91 (1994).
\item[{[19]}] C. D. Carone and H. Murayama, Phys. Rev. Lett. {\bf 74}, 3122 
(1995).
\item[{[20]}] P. B. Renton, Int. J. Mod. Phys. {\bf A12}, 4109 (1997).
\item[{[21]}] For example, MACRO had considered installing magnets to measure 
the charges of the observed leptons, but decided against it to reduce cost. 
G. Giacomelli, private communication.  Recently, a new experiment HANUL with 
such a capability has been proposed.  Y. Ho, private communication.
\end{enumerate}

\end{document}